\documentstyle[twocolumn,aps,aps]{revtex}
\begin{document}
\draft
 
\title{The effect of inter-edge Coulomb interactions on the transport 
between quantum Hall edge states}
 
\author{K. Moon} 

\address{
Department of Physics, University of California,
Davis, California 95616
}

\author{S. M. Girvin} 

\address{
Department of Physics, Indiana University,
Bloomington, Indiana 47405
}

\date{\today}
\maketitle
 
{\tightenlines
\begin{abstract}
In a recent experiment,
Milliken {\em et al.} demonstrated possible evidence 
for a Luttinger liquid
through measurements of the tunneling conductance between edge states
in the $\nu=1/3$
quantum Hall plateau. However, at low temperatures, a
discrepancy exists between the theoretical predictions
based on Luttinger liquid theory 
and experiment. We consider the possibility 
that this is due to long-range Coulomb 
interactions which become dominant at low temperatures.
Using renormalization group methods, we calculate the cross-over
behaviour from Luttinger liquid to the Coulomb interaction dominated
regime. The cross-over behaviour thus obtained seems to resolve
one of the discrepancies, yielding good agreement with experiment.
\end{abstract}
}
 
\pacs{PACS numbers: 72.10.-d, 72.10.Fk, 73.20.Dx, 72.15.Rn} 
 
\narrowtext
 
Recent progress in semiconductor nanostructure fabrication technology has 
ushered us into a world of reduced dimensionality which includes
one-dimensional 
quantum wires, two-dimensional electron systems like GaAs/AlGaAs 
heterostructures and so on. In addition, the discovery of high-$T_c$
superconductors, in which we believe the physics of correlated electrons
is at the origin of the many interesting and novel properties such as
non-Fermi liquid like normal phase behaviour, has increased our interest
in strongly correlated electron systems. 
Accordingly, the properties of one-dimensional interacting electron
systems have been studied in great detail\cite{Averin,Emery}. 
In one-dimension, inclusion of
short-range electron-electron interactions leads to the strongly correlated
electron systems known as Luttinger liquids\cite{Haldane}.
Recently, Kane and Fisher\cite{Kaneandfisher,Furusaki} have studied the 
transport
properties of a Luttinger liquid in the presence of a local impurity.
The physical realization of this system in terms of tunneling between
quantum Hall edge states is a particularly clean example of `boundary
critical phenomena.'

K. Moon {\em et al.}\cite{Kmoon}
have investigated the resonant tunneling between quantum Hall edge states
and numerically computed the universal resonance line shape,
obtaining results close to the exact result subsequently discovered by
Fendley {\em et al.}\cite{Fendley}.
These results show reasonable agreement with the experiment done by
Milliken {\em et al.}\cite{Milliken} provided that the data is 
normalized\cite{normalization} to
achieve the correct peak value $(1/3)(e^2/h)$. 
However, at very low temperatures, there is an additional discrepancy in
that the tunneling conductance falls below the theoretical
predictions.

Recently, Oreg and Finkel'stein\cite{Oreg} have considered the effect 
of the inter-edge Coulomb interaction on the propagation of the edge modes
in samples with constrictions.

In this paper, we study the effects of residual long-range Coulomb
interactions on the transport properties of a Luttinger liquid using a
renormalization group analysis and show that
these residual interactions may explain the
discrepancy between theory and experiment at low temperatures.

In the following, we assume that the bulk system exhibits the fractional 
quantum 
Hall effect with filling factor $\nu$ ($=1/n$, where $n$ is an odd integer). 
Since there is a charge excitation gap 
in the bulk, the low temperature physics is controlled by the 
{\em gapless} edge excitations\cite{Wen,Macdonald}. 
Consider a quantum Hall bar with width $w$ and length $L(> w)$ as 
illustrated 
in Fig.1.
In the quantum Hall bar geometry, electrons
at the two edges are moving in opposite directions.
The Lagrangian for the two corresponding edge excitations including the 
long-range Coulomb interactions\cite{Kmoon,Fisher} is given by
two-coupled chiral Luttinger liquids
\begin{eqnarray}
{\cal L} &=& \frac {1}{4\pi \nu} \int dx  \nonumber\\
&&\biggl\{ \partial_x\phi_R 
(i\partial_\tau\phi_R 
+v\partial_x\phi_R)
-\partial_x\phi_L (i \partial_
\tau\phi_L-v\partial_x\phi_L)\biggr\} \nonumber\\
&+&\frac {1}{8\pi^2 } \int dx dx^\prime \sum_{i=R,L} V_a(x-x')
\partial_x\phi_i \, 
\partial_{x'}\phi_i \nonumber\\ 
&+&\frac {1}{4\pi^2 } \int dx dx^\prime V_w(x-x')\partial_x\phi_R \, 
(-\partial_{x'}\phi_L) ,
\end{eqnarray}
where $v$ is the bare sound velocity of the edge wave and R(L) specifies 
right(left)-moving electrons.
The first term corresponds to the Lagrangian for pure chiral Luttinger liquids 
for the right and left-movers. The second and third terms describe
the intra- and inter-edge Coulomb interactions, 
respectively.
The operators $\phi_R$ and $\phi_L$ satisfy the following commutation
relations
\begin{equation}
[\phi_R(x),\phi_R(x^\prime)]=-[\phi_L(x),\phi_L(x^\prime)]
=-i\pi\nu\epsilon (x-x^\prime) ,
\end{equation}
where $\epsilon(x)=1 (-1)$ for $x>0 (<0)$ and
the intra- and inter-edge Coulomb interactions are given by
\begin{equation}
V_a(x)= \frac {e^2}{\epsilon \sqrt {x^2+a^2}} \; \; ,
V_w(x)= \frac {e^2}{\epsilon \sqrt {x^2+w^2}} 
\end{equation}
where $a$ is an ultraviolet cutoff on the scale of the mean particle spacing 
and $\epsilon$ is the dielectric constant.  
We define symmetric and antisymmetric modes
\begin{equation}
\theta \equiv \frac {1}{2 \sqrt \pi} (\phi_R - \phi_L) \; \; ,
\phi \equiv \frac {1}{2 \sqrt \pi} (\phi_R + \phi_L) 
\end{equation}
where the $\theta(x)$- and $\phi(x)$-fields satisfy the following algebra
\begin{equation}
[\theta(x),\phi(x^\prime)]=-i\frac {\nu}{2} \epsilon(x-x^\prime) .
\end{equation}
In the experiment, current is injected from source to drain as shown in Fig.1
and the voltage drop
$V_{SD}$ is measured, from which longitudinal conductance can be obtained.
The {\em net} current $J (=J_R-J_L)$ and charge operators 
$\rho (=\rho_R+\rho_L)$
are related to the $\phi_R$- and $\phi_L$-fields as follows
\begin{equation}
\rho=\frac {1}{2 \pi}(\partial_x\phi_R - \partial_x\phi_L)
=\frac {1}{\sqrt{\pi}} \partial_x \theta  ,
\end{equation}
\begin{equation}
J=\frac {1}{2 \pi}(\partial_x\phi_R + \partial_x\phi_L)
=\frac {1}{\sqrt{\pi}} \partial_x \phi.
\end{equation}
Following the above transformation and subtracting and adding $V_w(x)$,
we obtain the following Lagrangian
\begin{eqnarray}
{\cal L} &=& 
\frac {v}{2\nu}  \int dx \biggl((\partial_x\theta)^2+(\partial_x\phi)^2
\biggr)\nonumber\\
&+&\frac {1}{2\pi} \int dx dx' V_w\partial_x\theta \, \partial_{x'}
\theta\nonumber\\
& & +\frac {1}{4\pi} \int dx dx' (V_a-V_w)(\partial_x\theta \,
\partial_{x'}\theta+\partial_x\phi \, \partial_{x'}\phi).
\end{eqnarray}
Upon Fourier transformation and taking the long wavelength limit 
($k < w^{-1}$), one obtains the following Lagrangian
\begin{equation}
{\cal L} \cong \frac {v_R}{2\nu}   
\int dk \, k^2 \, [ \lambda(k) |\tilde{\theta}|^2 + |\tilde{\phi}|^2 ]
\end{equation}
where $v_R$ is a renormalized sound 
velocity given by
\begin{equation}
v_R=v \, \biggl[1 + \frac {\nu\alpha}{\pi\epsilon} \frac {c}{v}\ln 
(\frac {w}{a})
\biggr] ,
\end{equation}
$\alpha$ is the fine structure constant and $\lambda(k)$ is given by
\begin{equation}
\lambda(k)=1 + 2\chi\ln\biggl[\frac 
{2}{kw}\biggr] , 
\end{equation}
where we have defined the following dimensionless parameter $\chi$ which is 
a measure
of the strength of inter-edge Coulomb interactions
\begin{equation}
\chi\equiv \frac {\nu\alpha}{\pi\epsilon} \biggl(\frac {c}{v_R}\biggr) .
\end{equation}

Hence we obtain the generic Luttinger liquid Lagrangian with a renormalized
sound velocity $v_R$ plus the inter-edge Coulomb interaction term, which
gives $\lambda(k)$ a logarithmic divergence at small wavevectors.

Since $\partial_x \theta/\nu$ is canonically conjugate to the $\phi$-field 
from Eq.(5), 
it is convenient to integrate out the $\theta(x)$-fields and obtain the 
following Euclidean action for 
the $\phi(x)$-fields
\begin{equation}
{\cal S_{\em E}} \cong \frac {v_R}{2\nu} \sum_{\omega_n} \int dk \, 
\biggl\{ k^2 + 
\frac {1}{\lambda(k)}
\biggl( \frac {\omega_n}{v_R}\biggr)^2\biggr\} |\tilde{\phi}|^2 
\end{equation}
where $\omega_n=2n\pi/\beta$ is a Boson Matsubara frequency.
One can now directly read off the dispersion relation for the edge 
magnetoplasmon 
\begin{equation}
\hbar\omega=v_R k \sqrt{1+2\chi\ln\biggl[\frac 
{2}{kw}\biggr]} 
\label{eqm1}
\end{equation}
which agrees with the one obtained by M. Wassermeier 
{\em et al.}\cite{Wass}.
For small $k$, the linear dispersion relation, which is 
a characteristic of a Luttinger liquid, is modified to $k \sqrt{|\ln k|}$.

Now we are ready to study the transport through a narrow constriction
defined electrostatically by gates on both sides near the middle 
of a quantum Hall bar as shown in Fig.1. Since the right- and left-movers 
are spatially close together near the constriction, tunneling from one edge 
to the other can occur. Using a renormalization group analysis, 
Kane and Fisher 
\cite{Kaneandfisher} showed that 
for  weak back scattering, 
the effective barrier strength is renormalized to be
\begin{equation}
v_{\rm eff}\cong \frac {v_0}{T^{1-\nu}} .
\end{equation}
Hence, as the temperature is lowered, the effective barrier 
strength grows 
and the perturbative RG analysis based on the weak barrier limit 
breaks down. 
Furthermore the effects of Columb interactions 
become important in this regime. 
In this strong
barrier limit, the main contributions to the transport come from weak electron 
transmission as opposed to
the quasiparticle tunneling between edges which occurs in the weak 
barrier limit\cite{Kmoon,Wen}.
Since the electron creation operators $\psi_R^\dagger(x)$ and 
$ \psi_L^\dagger(x)$ 
are respectively given by 
$e^{i\phi_R(x)/\nu}$ and $ e^{-i\phi_L(x)/\nu}$, the 
electron transmission operator can be written 
$e^{i(\phi_L(x)+\phi_R(x))/\nu}\cong e^{i2\sqrt{\pi}\phi/\nu}$
\cite{Kmoon,Wen}.
Hence we have the following tunneling action $H_t$
(taking the tunneling to occur only at $x=0$)   
\begin{equation}
H_t=-t\int_0^\beta d\tau\cos (2\sqrt{\pi}\phi(\tau)/\nu).
\end{equation}
Using momentum-shell RG analysis, we obtain the following RG flow
\begin{equation}
\frac {dt}{d\ell}=\biggl[ 1 - \frac {2}{\nu\pi} 
\int_0^\infty dk \frac {\Lambda}
{{\Lambda^2/\lambda(k)} + k^2} \biggr] t
\end{equation}
with $\Lambda\cong e^{-\ell}$.
At finite temperature, the renormalization group flow is cut-off at
$\Lambda\cong T/a_T$ where $a_T$ is a dimensionless number of order unity. 
After integrating the above RG equation up to $\Lambda\cong T/a_T$, 
we obtain the
following effective (renormalized) tunneling amplitude $t_{\em eff}$ 
\begin{equation}
t_{\em eff}\cong \cases{ t_0 \frac {T_0}{T} \exp\biggl\{ -{1\over 3\nu\chi}
\biggl[ \biggl(1+2\chi {\rm ln}\frac {T_0}{T}\biggr)^{3/2} -1 
\biggr]\biggr\}&\cr 
t_0 \left(\frac {T_0}{T}\right)^{1-1/\nu}&\cr}
\end{equation}
where $T_0\equiv 2 a_T \hbar v_R/w$ is the cross-over temperature scale 
from Luttinger liquid
to the inter-edge interaction dominated regime, and the upper expression
is valid for $T\leq T_0$ while the lower applies for $T\geq T_0$. 
At low temperatures in the tail of the resonance,   %
the conductance $G_C^0$ can be obtained from the single 
particle 
hopping (or one-loop level) and is proportional to $t_{\em eff}^2$
\begin{equation}
G_C^0\cong t_0^2 \biggl(\frac {T_0}{T}\biggr)^2 \exp\biggl\{ -{2\over 3\nu
\chi}
\biggl[ \biggl(1+2\chi {\rm ln}\frac {T_0}{T}\biggr)^{3/2} -1 \biggr] 
\biggr\}.
\label{GOC}
\end{equation}
In the absence of inter-edge Coulomb interactions (i.e. $\chi\rightarrow 0$),
we recover the familiar Luttinger liquid behaviour\cite{Kaneandfisher}
\begin{equation}
G_L^0\equiv G_C^0(\chi\rightarrow 0)\cong t_0^2 
\biggl(\frac {T_0}{T}\biggr)^{2(1-1/\nu)} .
\label{GOL}
\end{equation}
In the extreme Coulomb interaction dominated regime ($\chi\gg 1$),  %
Eq.(\ref{GOC}) reduces to %
the following result
\begin{equation}
G_C^0(\chi\gg 1)\cong t_0^2 \biggl(\frac {T_0}{T}\biggr)^2 
\exp\biggl\{ -{2\over 3\nu}
\sqrt{2\chi} \biggl({\rm ln}\frac {T_0}{T}\biggr)^{3/2}\biggr\}
\end{equation}
which L.I. Glazman {\em et al.}\cite{Glazman} obtained by studying
the transport properties of a
charge density wave
through a single barrier.
The underlying physics can be understood as follows. 
We know that the Luttinger liquid with $\nu<1$ has a charge density wave (CDW)
correlation function which decays algebraically with distance.
In the presence of the Coulomb interaction, the CDW correlation
function decays slower than any power law\cite{Schulz}.
Since a CDW is easily pinned by even a single impurity, one can expect  
the conductance to be further reduced in the presence of Coulomb interactions.

The universal resonance line shape $G_L$
for the Luttinger liquid with short range interactions at $\nu=1/3$ 
was obtained previously by K. Moon {\em et al.}\cite{Kmoon} using Monte carlo
methods and subsequently the exact solution
was obtained by Fendley {\em et al.}\cite{Fendley}.
Using Eq.(\ref{GOL}) and Eq.(\ref{GOC})  %
one can calculate the tunneling conductance in the presence of Coulomb
interactions 
by taking into account the leading corrections in the following 
approximate  %
way
\begin{equation}
G_\chi (T)\cong \cases{ G_L \, (G_C^0/G_L^0)  &for $T\leq T_0$ \cr
                  G_L                    &for $T\geq T_0$. \cr}
\label{eq22}
\end{equation}
The conductance $G_\chi(T)$ thus obtained has {\em no} explicit dependence on 
the bare tunneling amplitude $t_0$ and satisfies the following properties.
When the tunneling amplitude is very small, $G_L/G_L^0$ approaches unity 
and the conductance 
is given by $G_C^0$ as expected.
At relatively high temperatures, since the thermal correlation length 
($\xi_\beta\cong v\hbar/2\pi T\nu$) 
is rather short and the long-range
Coulomb interaction is less effective,
$G_C^0/G_L^0$ becomes $1$ and the conductance is given by 
the Luttinger liquid value $G_L$. 
We want to emphasize that there exist two crossover scales\cite{KaneP}.
In addition to the scale $T_0$, describing the onset of Coulomb interaction,
there is also the scale $T_1$ ($\propto v_0^{1/(1-\nu)}$), which governs 
the crossover from the weak to
the strong barrier regime.
Following Eq. (\ref{eq22}), for $T\leq T_0\leq T_1$, the conductance can be 
written    
\begin{equation}
\tilde{G}_\chi \biggl[\frac {T_1}{T},\frac{T_0}{T}\biggr] = 
G_L\biggl[\frac {T_1}{T}\biggr] 
F_\chi\biggl[\frac {T_0}{T}\biggr]
\label{eq23}
\end{equation}
where $G_L$ is the Luttinger liquid contribution and the function $F_\chi$ 
is given by
\begin{equation}
F_\chi(x) = x^{2/\nu} 
\exp\biggl\{ -{2\over 3\nu\chi}
\biggl[ \biggl(1+2\chi {\rm ln}x\biggr)^{3/2} -1 \biggr]
\biggr\}.
\label{decompose}
\end{equation}
One can clearly see in Eq.(\ref{eq23})
the breakdown of single-variable scaling 
due to the existence of the additional Coulomb scale $T_0$.

Now we want to compare our results with the recent experiment by 
Milliken {\em et al.}
\cite{Milliken}.
The experiment is performed on a 
GaAs/AlGaAs heterostructure. A point contact gate voltage applied in the 
middle of the sample pinches off a narrow channel. The bulk system is in 
the $\nu=1/3$
fractional quantum Hall regime. Fine tuning of the gate voltage 
produced several resonance peaks and they measured the temperature dependence 
of the tunnelig conductance.
The nominal size of the quantum Hall bar is roughly 
$60\times 60 \mu m^2$ and
dielectric constant $\epsilon\cong 13$.

The mean electron density $n$ is about $1.0\times 10^{11} cm^{-2}$.
Hence the mean particle spacing $a$ is approximately $300\AA$ and the 
edge wave velocity $v_R$ is about $10^5 {\rm m/sec}$
\cite{Allan,Laux}.
Using the parameters above, fixes $\chi\sim 0.18$. 
In Fig.2, the experimental data obtained by 
Milliken {\em et al.}\cite{Milliken}
are compared to our results showing nice agreement with
a best fit value for the cross-over temperature $T_0$ of
about 60 mK which gives $a_T \approx 2.5$ close to unity as expected  
and $T_1$ is chosen to be 60 mK.  
The somewhat poorer data collapse in the tail of the scaling curve
at low temperatures
might be a manifestation of the breakdown of scaling in terms of a single
variable due to Coulomb interactions.

Unfortunately, we can not know for certain at this point how significant
this good fit (using essentially one additional parameter)
really is.  While the nominal sample geometry is a $60\mu$
square, the actual edge geometry under the near pinch-off conditions of
the experiment is
unknown.  At the cross over temperature, the thermal correlation length
is only about $15\mu$ so that the asymptotic long-distance approximation
($k < w^{-1}$) is not likely to be valid.  
In addition, the sample has a back gate, which presumably screens out the 
long-range piece of Coulomb interaction leading to the dipole-like interaction
($\propto 1/r^3$)\cite{Fisher,Milliken}.

Nevertheless, the fact that a good fit
is obtained with physically reasonable values of the parameters lends
credence to the idea that long range Coulomb forces are playing a
significant role at the lowest temperatures.  
Clearly however, more
experimental studies with well-defined geometries
are needed to address these issues.

In conclusion, we have approximately computed the cross-over 
from the Luttinger liquid to the 
inter-edge Coulomb interaction dominated regime  of a quantum Hall bar
with back scattering at a narrow constriction.  It is clear that further
experiments on narrower samples with well-characterized geometries
would be very useful in probing this interesting phenomenon.

\acknowledgements{
It is a pleasure to thank M.P.A. Fisher, C.L. Kane, A.H. MacDonald, A. Ludwig, 
and P. Fendley 
for illuminating discussions. We thank F.P. Milliken for providing us with 
their experimental data and information on sample geometry.

The work at Indiana was supported by the NSF through Grant No. DMR-9416906.
The work at UC-Davis was supported by NSF Grant No. DMR 92-06023 and by the
Los Alamos National Laboratory through a LACOR grant.
}

\begin{figure}
\caption{Schematic diagram for a quantum Hall bar of width $w$ and length
$L$. The gates $G$ are placed in the middle of the Hall bar to make a narrow 
constriction. Charges can tunnel through the barrier {\em or} be
reflected. The source $S$ and drain $D$ are located at each end of the bar.
}
\label{fig1}
\end{figure}

\begin{figure}
\caption{The ($+$) symbols correspond to the experimental data obtained 
by Milliken {\em et al.} and the open circles are from the Monte Carlo data
by K. Moon {\em et al.}  
based on the Luttinger liquid theory. The solid curves are obtained
by taking $\chi\approx 0.18$ and $T_1$ ($\sim T_0$) is chosen to be 
about 60 mK. 
}
\label{fig2}
\end{figure}


\begin{references}
\bibitem{normalization}  This normalization  problem 
is not understood at present.  The experimental conductance never
achieves the universal peak value $(1/3)(e^2/h)$ but is considerably
smaller.  In order to actually hit the resonance peak both the real and
imaginary parts of the backscattering amplitude have to be tuned to zero
simultaneously.   This requires two adjustable experimental parameters
and may not have been achievable in the existing experiment.
\end{references}
\end{document}